\documentclass{aa}
\input epsf

\usepackage{graphics}
\newcommand{\be}{\begin{equation}}
\newcommand{\ee}{\end{equation}}
\newcommand{\bea}{\begin{eqnarray}}
\newcommand{\eea}{\end{eqnarray}}

\def\lta{\ifmmode {\,\mathbin{\lower 3pt\hbox   
    {$\,\rlap{\raise 5pt\hbox{$\char'074$}}\mathchar"7218\,$}}}
    \else {${\mathbin{\lower 3pt\hbox
    {$\rlap{\raise 5pt\hbox{$\char'074$}}\mathchar"7218\,$}}}
    $}\fi}
\def\gta{\ifmmode {\mathbin{\lower 3pt\hbox   
    {$\,\rlap{\raise 5pt\hbox{$\char'076$}}\mathchar"7218\,$}}}
    \else {${\mathbin{\lower 3pt\hbox
    {$\rlap{\raise 5pt\hbox{$\char'076$}}\mathchar"7218\,$}}}
    $}\fi}

\begin{document}


\title{Limits on the Density of 
Compact Objects from High Redshift Supernovae}

\author{U. Seljak\inst{1}
\and
D.E. Holz\inst{2}
}

\institute{Dept. of Physics, Princeton University, Princeton, NJ 08544, USA
\and
Albert-Einstein-Institut, Max-Planck-Institut f\"ur Gravitationsphysik,
Am M\"uhlenberg 5, 14476 Golm, Germany 
}

\date{\today / Submitted to {\it A \& A Letters}}

\titlerunning{Limits on the Density ...}

\maketitle

\begin{abstract}
Due to the effects of gravitational lensing,
the magnification distribution of high redshift supenovae can be 
a powerful discriminator between smooth dark matter and 
dark matter consisting of compact objects. 
We use high resolution N-body simulations in combination 
with the results of simulations with
compact objects to determine 
the magnification distribution for a Universe with an arbitrary 
fraction of the dark matter in compact objects. Using these 
distributions we determine the number of type Ia SNe required to measure
the fraction of matter in compact objects.
It is possible to determine a 20\% fraction of matter in
compact objects with
100-400 well measured SNe at $z \sim 1$, assuming the background
cosmological model is well determined.

%
\end{abstract}

\section{Introduction}
Current cosmological data indicates that the density of matter 
in the Universe  
is around $\Omega_{\rm m} \sim 0.3$ (in units of critical density). 
Of this, big bang nuclesynthesis requires 15--20\% be in the form of
baryons, with the rest in some other more exotic form. Even among the 
baryons only $\Omega_* \sim 0.01$ has been accounted for directly. The 
rest could be in warm gas (Cen \& Ostriker 1999), or in compact objects 
such as brown dwarfs, white dwarfs, or neutron stars. 
The nonbaryonic dark matter could be composed of a smooth
microscopic component, such as axions, or alternatively it may
be composed of compact objects,
such as primordial black holes (Crawford and Schramm 1982). 
Although compact objects have been searched 
for in our own halo using microlensing studies 
(see Sutherland 1999 for a review), 
results are as of yet inconclusive. While the most straightforward 
interpretation is that a large
fraction of the halo (up to 100\%) is composed 
of compact objects of roughly 0.5M$_{\sun}$, 
other scenarios using existing stellar 
populations remain viable. 
The bottom line is that,
at present, the form of the bulk of the dark matter remains unknown.

In recent years there have been tremendous advances in our
ability to observe and characterize type Ia
supernovae. In addition to dramatically increasing the
number of SNe observed at high redshift, the intrinsic peak
brightness of these supernovae is now thought to be known to
within about 15\%. These supernovae are thus excellent
standard candles, and by measuring the spread in their
observed brightnesses it may be possible to determine the
nature of the lensing, and thereby infer the distribution of
the lensing matter.

It has long been recognized that the lensing of SNe can be
used to search for the presence of compact objects in the
Universe (Linder, Schneider \& Wagoner 1988; Rauch 1991).
Since the amount of matter near or in the beam determines
the amount of magnification of the image, the magnification
distribution from many SNe can probe for the presence of
compact objects in the Universe.  If the Universe consists
of compact objects, then on very small scales most of the
light beams do not intersect any matter along the line of
sight, resulting in a dimming of the image with respect to
the filled-beam (standard Robertson-Walker) result.  On
occasion a beam comes very near a compact object, resulting
in a tremendous brightening of the ensuing image.  In such a
Universe the magnification distribution will be sharply
peaked at the empty beam value, and will possess a long tail
towards large magnifications. The lensing is sensitive to
objects with Einstein rings larger than the linear extent of
the SNe (roughly $10^{15}$cm at their maximum, which gives a
lower limit on the mass of the lenses of $10^{-2}$M$_{\sun}$).

Lensing due to compact objects
is not, however, the only way to modify the flux of a SN.
Even smooth microscopic matter, such as the lightest SUSY particle or axions, 
are expected to clump on large scales.
The effect on the magnification distribution 
depends on the clumpiness of the Universe.
If the clumping of matter in the
Universe is very nonlinear then all of the matter will reside 
in dense halos, and the filaments connecting them, and 
there will be large empty voids 
extending tens or even 
hundreds of megaparsecs in diameter. There will thus
be a large probability that a given 
line of sight will be completely devoid of matter, and so will give a large
demagnification (as compared to the pure Robertson-Walker result).
A simple way to estimate 
the importance of this effect is to
compare the rms scatter in magnification to the demagnification 
of an empty beam relative to the mean (given by the filled
beam value). 
Both can be calculated analytically, 
the former as an integral over the nonlinear power spectrum, and the 
latter as an integral over the combination of distances. 
The result depends on the redshift of interest, but for most realistic 
models the rms is smaller than the empty beam value at $z
\gta 1$.
This means that, at least qualitatively, it should be possible 
to distinguish between compact objects and smoothly distributed 
matter at such redshifts.

In this letter we extend previous work in several aspects. 
First,
we provide the formalism to investigate models with a combination of
both compact 
objects and smooth dark matter. 
Although Rauch (1991) and 
Metcalf and Silk (1999) have explored the use of lensing of supernovae
to detect compact objects, they concern themselves with
distinguishing between two extreme cases:
either all or none of the matter in compact objects.
Our formalism allows us to address the more general question 
of how well the fraction of dark matter in compact objects can
be {\em measured} with any given SN survey.
Both baryonic and dark matter are dynamically significant
and could contribute to the lensing signal.
For example, we can imagine four simplistic scenarios, in which the
baryonic and dark matter are each in one of two states:
smoothly distributed or clumped into compact objects (with masses
above $10^{-2}$M$_{\sun}$). To distinguish among these cases we need 
to be able to differentiate between 0\%, 20\%, 80\%, and
100\% of the matter in compact objects.

Second, we use
realistic cosmological N-body simulations (Jain, Seljak and White 1999) 
to provide the 
distribution of magnification for the smooth microscopic component. 
Previous works (Metcalf and Silk 1999; Holz and Wald 1998) make the
simplifying assumption of an 
uncorrelated distribution of halos to determine this distribution.
As shown in Jain et al.
(1999; JSW99), there are differences in the probability 
distribution function (pdf) of magnification for models with  
different shapes of the power spectrum 
and/or different values of cosmological parameters. For example, 
open models exhibit large voids up to $z \sim 1$, and their pdf's
extend almost to the empty beam limit. On the other hand, flat 
$\Omega_{\rm m}$ models have less power (being normalized to the 
same cluster abundance), and are more linear at higher $z$, resulting 
in a more Gaussian pdf. 
These differences become particularly important when we consider models
in which only a fraction of the total matter is in compact objects.
Magnification
distributions in such cases differ only weakly from the smooth matter 
case, and an accurate description of the pdf is of particular importance.
Finally, 
we also discuss the cross-correlation of SN magnification with convergence
reconstructed from shear or magnification in the same field. As
will be presented in the discussion section,
this can be used as a further probe of compact objects.

\section{Magnification probability distribution}
We wish to derive the magnification probability distribution
function (pdf)
in a Universe with both compact objects and smooth dark matter. 
We begin with the magnification pdf for smoothly distributed
matter, $p_{\rm{LSS}}(\mu,z)$,
since this background distribution is present in all cases.
For convenience we define the magnification, $\mu$,
to be zero at the empty beam value.
We use the pdf's computed from N-body simulations in JSW99,
obtained by counting the number of pixels in a 
map that fall within a given magnification bin. 
We explicitly include the dependence of the pdf on the
redshift of the source.
The two main trends with increasing $z$ are
an increase in the rms magnification, and an
increasing Gaussianity of the pdf
(see figure 15 in JSW99). 
As we increase $z$ 
we are superimposing more independent regions
along the line of sight, and the resulting pdf approaches
a Gaussian by the central limit theorem.

A possible source of concern is that the resolution
limitations of the N-body simulations might corrupt the
derived pdf's in ways which crucially impact our results.
In principle we would like to resolve all scales down to 
the scale of the SN emission region ($10^{15}$cm 
in linear size). Fortunately such high resolution is unnecessary,
as there is very little power in the matter correlation
function on such small scales. 
The contribution to the second moment of the magnification
peaks at an angular scale of
$\theta>3'$ (see figure 8 in JSW99);
scales smaller than this do not significantly change the 
value of the second moment.
These smaller scales, however, are relevant for
the high magnification tail of the distribution, and even the largest 
N-body simulations are resolution limited in the centers of
halos.
As high magnification events are very rare in the small SN
samples being considered, limitations in the resolution of
the high magnification tail of the distribution are not of
great concern.
The simulations are very robust around the peak 
of the magnification pdf, with lower
resolution PM simulations giving results in good agreement with 
higher resolution $P^3M$ simulations (figure 20 in JSW99). 
As these $P^3M$
simulations converge for the 2nd moment 
we may conclude that, aside from the high-magnification
tail, the pdf's obtained from these simulations do not
suffer from the limitations of finite numerical 
resolution. This is also confirmed by smoothing the map by a factor of two, 
and comparing the pdf's before and after the smoothing.
The resulting pdf's are
very similar for all models, indicating that small scale power
has little effect on the region of most interest, near the peak of the pdf.

\begin{figure}
\epsfxsize=3.5in \epsfbox{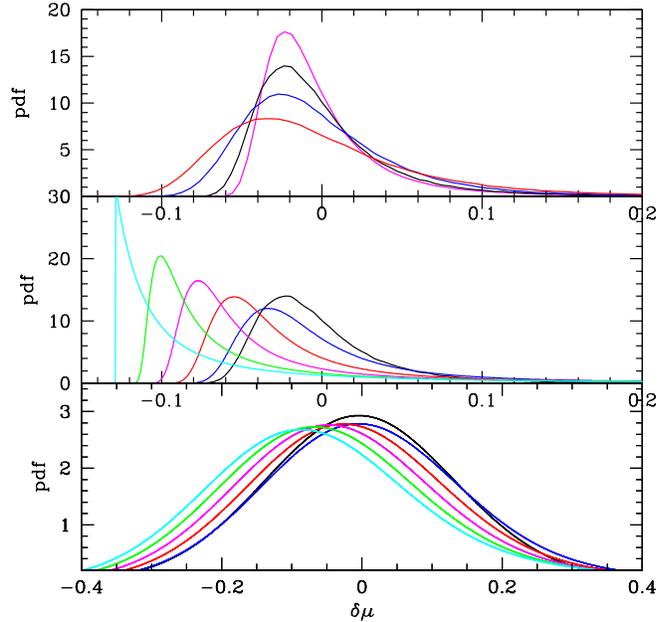} 
\caption{One point distribution function of 
magnification relative to the mean. The top panel
shows the pdf's in the absence of compact objects for 
(from top 
to bottom)
OCDM (open), $\lambda$CDM, $\tau$CDM and standard CDM
Universes (see JSW99 for a detailed description of the models).
In the middle panel the pdf's are given 
as a function of the fraction of total matter in compact objects, $\alpha$,
for the $\lambda$CDM model.
The curves from right to left are for
$\alpha=0,0.2,0.4,0.6,0.8,1$.
The bottom panel gives the same pdf's as the middle one,
convolved with the (intrinsic and observational)
``noise'' of the SNe ($\sigma_{\mu}=0.14$).
}\label{fig2}
\end{figure}

The pdf's are shown in figure \ref{fig2}, plotted against deviations
from the mean magnification,  $\delta \mu =\mu-\bar{\mu}$
($\delta\mu = 0$ corresponds to the mean (filled beam) value).
The mean magnification, $\bar{\mu}$, 
is given by the difference
between the empty 
beam and the mean (filled) beam values, which for SN
at $z=1$ is $\bar{\mu}=0.24$ for $\Omega_{\rm m}=1$, 
$\Omega_{\lambda}=0$ models such as 
standard CDM model (with $\Omega_{\rm m} h=0.5$) 
or $\tau$CDM model (with $\Omega_{\rm m} h=0.21$),
$\bar{\mu}=0.13$ for $\lambda$CDM model with
$\Omega_{\rm m}=0.3$, $\Omega_{\lambda}=0.7$, and $\bar{\mu}=0.09$ for open (OCDM) 
model with
$\Omega_{\rm m}=0.3$,
$\Omega_{\lambda}=0$.

In contrast, in a Universe filled with a uniform comoving density of
compact objects the pdf depends on a single parameter, 
the mean magnification $\bar{\mu}$, or equivalently,
the mean convergence $\sigma$. The two are related via
$\bar{\mu}=(1-\sigma)^{-2}-1$ ($\bar{\mu}\sim 2\sigma$ if $\sigma \ll 1$). 
The pdf rises sharply from $\mu=0$, and 
drops off as $\mu^{-3}$ for high $\mu$ (Paczy\' nski 1986). Based on 
Monte-Carlo simulations, Rauch (1991) gives a fitting formula for the 
pdf:
\begin{equation}
p_{\rm C}(\mu,\sigma)=2\sigma_{\rm {eff}}\left[ {1-e^{-b\mu} \over 
(1+\mu)^2-1}\right]^{3/2},
\label{pdfc}
\end{equation}
where $b=247e^{-22.3\sigma}$ and
$\sigma_{\rm {eff}}$ is chosen so that the pdf integrates to unity.
Note that this expression is only valid for $\sigma<0.1$, and can only be 
used for SN with $z < 1$--2, depending on the cosmology. 

To combine the two distributions we consider a model where
a fraction $\alpha$ of the matter is in compact objects, and where
these compact objects trace the underlying matter distribution.
Suppose a given line of sight has magnification $\mu$ in the absence of 
compact objects. In the presence of compact objects the mean magnification 
along this line of sight remains unchanged. Since
the smooth component contributes a SN magnification of $(1-\alpha)\mu$,
the effect of compact objects is described with 
a pdf that gives a mean magnification of $\alpha\mu$.
The combined pdf, $p(\mu)$,
is given by integrating over the whole distribution,
\begin{equation}
p(\mu;\alpha,z)=\int_{0}^{{\mu \over 1-\alpha}}
p_{\rm {LSS}}(\mu',z)p_{\rm C}[\mu-\mu'(1-\alpha),\alpha\mu'/2]d\mu'.
\label{pdft}
\end{equation}

The middle panel in 
figure \ref{fig2} shows the magnification pdf for a range of values
of $\alpha$,
for a cosmological model with $\Omega_{\rm m}=0.3$, $\Omega_{\lambda}=0.7$, and 
$\sigma_8=0.9$. 
The larger the value of $\alpha$, the closer
the peak of the distribution to the empty beam value.
As $\alpha$ increases from zero the pdf becomes wider, as the
compact objects increase the large magnification tail.
As $\alpha$ increases beyond $\alpha=0.2$, however,
the distribution begins to narrow, since more and more lines
of sight are empty and thus closer to the empty beam value. 
Note in particular the similarity between the $\alpha=0.2$
pdf and the
$\tau$CDM model with $\alpha=0$ (upper panel in figure \ref{fig2}).

We need to further convolve these
distributions with the measurement noise and the scatter in the intrinsic 
SN luminosities. Current estimates for these is 
0.07 magnitudes for rms observational noise and 0.12
magnitudes for intrinsic scatter.
The two combined give an additional rms scatter in magnification of 0.14
(Hamuy et al. 1996).
To model this noise we convolve all the pdf's 
with a Gaussian of width $0.14$. The resulting pdf's are shown in the
bottom panel of figure \ref{fig2}. The distinction between the different values of 
$\alpha$, although small, is still apparent.
In the next section we calculate how many SNe are required
to distinguish between the different curves in the bottom panel, and thereby
measure $\alpha$.

\section{Maximum Likelihood Analysis}
We assume that the SNe are independent events.
The likelihood function for the combined sample is then just the 
product of the individual SN likelihood functions,
\begin{equation}
L(\alpha)=\Pi_i p(\mu_i;\alpha,z_i).
\end{equation}
Redshift information could
serve as an important confirmation that the effects observed
are due to gravitational lensing,
since the shapes and positions of the pdf's 
evolve with redshift in a known manner.
This redshift dependence does not, however,
significantly increase the statistical power of the determination, compared to 
a sample where all of the SNe are at the mean redshift
of the sample. To simplify the analysis we 
therefore assume that all of the SN are at a fixed redshift, z=1,
and drop the 
redshift dependence of $p$. Taking the log and ensemble averaging we find
\begin{eqnarray}
\langle \ln L(\alpha) \rangle &=&\Big\langle \sum_i \ln p(\mu_i;\alpha) 
\Big\rangle \nonumber \\
&=& N_{\rm SN}\int p(\mu;\alpha_0)\ln p(\mu;\alpha)d\mu,
\end{eqnarray}
where $N_{\rm SN}$ is the number of observed SNe and 
$\alpha_0$ is the assumed true value of $\alpha$.

An estimate of the unknown parameter $\alpha$ is given by maximizing the 
log-likelihood
function. The ensemble average of this gives 
the solution $\alpha=\alpha_0$ (i.e. the estimator is asymptotically unbiased).
The error on the determination of the
parameter $\alpha$ is given by the curvature of the 
negative log-likelihood function around its maximum.
The ensemble average of this minimum variance is
\begin{eqnarray}
\sigma_{\alpha}^{-2}&=&
\Big\langle \Big[{\partial  \ln L(\alpha) \over \partial \alpha} \Big]^2\Big\rangle
\nonumber \\
&=&N_{\rm SN}\int p(\mu;\alpha) 
\Big[{\partial \ln p(\mu;\alpha) \over \partial \alpha}\Big]^2d\mu
,
\label{sigalpha}
\end{eqnarray}
which is to be evaluated at $\alpha=\alpha_0$.
According to the Cram\' er-Rao theorem, $\sigma_{\alpha}$ gives
the smallest attainable error on $\alpha$ for an unbiased estimator. 
As expected,
the error decreases as the inverse square root of the number of SNe.
Equation~\ref{sigalpha} determines the number of SNe
required to achieve a given level of confidence in the
measurement of $\alpha$, given a true value $\alpha=\alpha_0$.
The case of $\alpha_0=0$ 
gives the sensitivity in the case of a Universe with a small
fraction of matter in compact objects.
In this case, for $\lambda$CDM model
we find 
$\sigma_{\alpha} \sim 
1.4 N_{\rm SN}^{-1/2}$. This includes information from the full pdf, so 
any large differences between the models in the tail of the pdf would 
be statistically significant.
Although this tail is susceptible to systematic effects 
generated by a lack of knowledge of the pdf, it
will not be probed by small numbers of SNe. 
To exclude the tails we redid the integral in equation \ref{sigalpha}
including only the information within $\pm 2\sigma_{\mu}$ around the mean.
The resulting error increases to $\sigma_{\alpha} \sim
1.8 N_{\rm SN}^{-1/2}$.
This means that we need on the order of
100 SNe to determine $\alpha$ to 20\%,
and around 1000 to determine 
$\alpha$ to 5\%, all with one-sigma errors and assuming $\alpha_0=0$. 
The variance gradually increases with
$\alpha_0$, and at $\alpha_0=1$ the required number 
of SN increases by a factor of 2.

The variance in $\alpha$ scales 
roughly linearly with the rms noise variance $\sigma_{\mu}$, 
so if the scatter in SN is larger
the corresponding error in $\alpha$ increases.
Variance also scales roughly inversely with the separation 
between mean and empty beam $\bar{\mu}$. For an open Universe, $\bar{\mu}$
is $7/10$ the value in the flat ($\Omega_{\rm m}=0.3$) Universe,
so one needs roughly twice the number of SNe
to reach the same sensitivity. 
For a
flat Universe with $\Omega_{\rm m}=1$ and $\bar{\mu}=0.24$ 
the statistical significance 
increases, and one needs one quarter the SNe for the same sensitivity.

One can also test for the systematic effects introduced by 
using an incorrect model for the smooth component. 
To do this we did an analysis of a $\lambda$CDM Universe,
``mistakenly'' assuming it was an OCDM one.
This 
results in a bias on the parameter $\alpha$
which can be as high as 20\%, so that a precision test with this method is 
only possible once the parameters of the 
cosmological model are known precisely.

\section{Discussion}
Results obtained in the previous section indicate that
high-$z$ SNe can enable an accurate 
determination of the fraction of matter in compact objects.
If the estimates of the errors used here are reasonable,
around 100 SNe at $z=1$ are required to make a 20\% determination
of the fraction of dark matter in compact objects.
Although such numbers of SNe are not currently available, 
they can be expected in the near future as the high-redshift
supernovae surveys continue their
observations.\footnote{In addition, SNAPSAT (Perlmutter 1999),
a proposed satellite dedicated to observing SNe,
would easily meet the requirements, with ${}\sim2000$ high-$z$ SNe per year.}
Our results are in broad agreement with the results of Metcalf 
and Silk (1999), who find that about 50-100 SN are 
required to distinguish $\alpha=0$ from $\alpha=1$ with $90\%$
confidence.
An important limitation will be systematic effects caused by
uncertainties in the magnification distributions.
For the smooth component
these can be obtained using high resolution N-body 
simulations. We have argued that our results are reliable in the 
regime of application and can in any case be verified 
with higher resolution simulations in the future. 
Lack of knowledge of the underlying cosmological model may also 
introduce systematic effects, but again this can be expected to be
better determined in the future.  

Another important systematic error is our ignorance of the
intrinsic dispersion in brightness of the observed SNe. Our
simple assumption of a Gaussian noise profile in the
measurement of the peak luminosities of SN Ia's is certain to
break down at some level. This noise estimate is based upon
phenomenological observations of SNe, and needs to be borne
out both by further observations (especially at low
redshift, where lensing effects are negligible and
independent calibrations are available) and by theoretical
models.  Deviations from Gaussian noise are most likely to
be important in the tail of the magnification pdf's, which can be
excluded in the analysis.  The lack of a precise knowledge
of the intrinsic peak values of the SNe is less likely to
cause a shift in the peak of the magnification pdf's, which
is the main signature of the lensing effect.

A further source of concern is redshift evolution of the
intrinsic properties of the SNe. It is possible that both
the mean and the higher moments of the distribution of peak
luminosities of type Ia SNe varies with redshift, and this
could pose significant challenges to an accurate measurement
of lensing effects. Improvements in the determination of
such effects will occur as the size of the data sets at both
low and high redshifts are increased, and direct comparisons
of observations are available. An important consistency
check will be to demonstrate that the shift of the peak of
the distribution as a function of redshift is consistent
with theoretical expectations (under the assumed value of
$\alpha$).

A complimentary method to obtain the pdf due to the 
background smooth matter component is to use weak lensing 
observations. This would provide the pdf 
directly from data, avoiding entirely 
the need for cosmological simulations. As
discussed in \S 2 most of the power is on scales larger than 
3', so the pdf will almost converge to the correct one
even if one smoothes the beam 
at this scale. Weak lensing surveys can provide maps of 
the magnifications by reconstructing the projected mass density
from the shear
extracted using galaxy ellipticities. The distribution of 
magnifications gives a pdf convolved with the random noise from 
galaxy ellipticities. Given an rms ellipticity of $0.4$, we find 
that about 50 galaxies per square arcminute 
are required to give an rms noise comparable to 
the noise in the SNe. This number is similar to the density of 
galaxies at deep ($m_I \sim 26$) exposures. Such
galaxies have a mean redshift ($z \sim 1$) comparable to the
SNe under discussion. We can therefore 
choose the size of a patch such that its rms noise
agrees with the noise in the SN data. Such a pdf can then
be directly compared to that from a SN sample (provided the differences
in the redshift distribution are not too large), and  
any differences between the weak lensing and SN pdf's would indicate
the presence of compact objects or some other source of small scale 
fluctuation in magnification.
Note that one 
cannot test for compact objects by simply using cross-correlation 
of the two magnifications, as compact objects do not change 
the mean magnification in a given line 
of sight, and thus the cross-correlation coefficient
remains unchanged. The increased scatter, however, could
provide the desired signature.

In conclusion, the magnification distribution from several hundred 
high redshift type Ia SNe has the 
statistical power to make a 10-20\% determination
of the fraction of the dark
matter in compact objects.
It remains to be seen whether this statistical power can be 
exploited at its maximum, or whether systematic effects will
prove to be too daunting. 

\begin{acknowledgements}
US ackowledges the support of NASA grant NAG5-8084 and 
B. Jain and S. White for collaboration on JSW99.
\end{acknowledgements}

\end{document}